# A Lightweight and Post-Quantum Secure Framework for IEC 61869-9 Sampled Value Communication

S.M. Suhail Hussain, *Senior Member, IEEE*, Arman Ahmad, Mohammad Tayyab, *Senior Member, IEEE*, and Shaik Mullapathi Farooq, *Senior Member, IEEE*

***Abstract*— Securing IEC 61869-9 Sampled Values (SV) is challenging because process-bus communication must satisfy stringent real-time constraints while supporting standardized high-rate publication profiles. This paper presents an experimentally validated security framework that combines lightweight per-frame authentication for operational SV traffic with post-quantum-capable key establishment protocol. For message integrity, the proposed method applies field-selective authentication employing optimized Chaskey-12 to reduce per-packet computational overhead. For trust establishment, the paper introduces an ML-KEM-based pairwise authentication and key-establishment procedure. The pairwise protocol is analyzed in the Quantum Random Oracle Model and is also verified with AVISPA tool under the Dolev-Yao adversarial model. A C-based publisher/subscriber prototype is implemented on a two-node process-bus testbed. Performance is evaluated across the eight IEC 61869-9 SV packet profiles using HMAC-256, AES-GMAC-128, Blake-2s, Chaskey-12, and a compiler-optimized Chaskey-12 implementations. These results indicate that optimized Chaskey-12 achieves ~90% lower latency than HMAC on SV packets. The proposed security framework is a practical and scalable candidate for protecting IEC 61869-9 SV traffic on resource-constrained digital-substation devices.**

***Index Terms*— IEC 61850, IEC 61869-9, Sampled Values, IEC 62351-9, Chaskey-12, ML-KEM, post-quantum cryptography.**

## I. INTRODUCTION

THE modern substation automation system (SAS) has evolved from conventional hardwired installations toward Ethernet-based digital process bus architectures in which interoperability, modularity, and high speed data exchange are central design goals. The feasibility of this transition has already been demonstrated through IEC 61850-9-2 process bus protection implementations that achieved performance comparable to conventional hardwired arrangements when the communication network was properly engineered [1].

The UCA International Users Group guideline for IEC 61850-9-2 Light Edition (9-2 LE) served as the de facto interoperability profile for practical process-bus deployment by defining a realizable subset for streaming synchronized current and voltage measurements from merging units to multiple subscribing intelligent electronic devices (IEDs) [2]. This evolution was subsequently consolidated by IEC 61869-9, which specifies the digital interface for instrument transformers and standardizes different SV publication profiles for SAS applications, where the sampling rate and the number of ASDUs packed per frame jointly determine the packet rate and the available interarrival time at the subscriber [3]. Because these measurements directly drive protection and control decisions, the cyber risk is consequential; false SV injection has been shown to disrupt protection functions in IEC 61850 automated substations [4]. Therefore, any security mechanism for IEC 61869-9 SV must preserve message authenticity and integrity while respecting the stringent timing constraints imposed by standardized high-rate streams [3]. While much of the IEC 61850 cybersecurity literature has concentrated on GOOSE messaging [5], [6], SV is equally important because it carries the continuous measurements that directly drive protection and control decisions. Moreover, the very high SV publication rates standardized in IEC 61869-9 impose extremely tight per-frame processing budgets and stricter implementation constraints, including memory usage, especially when an IED subscribes to multiple concurrent streams [7].

Beyond cryptographic protection, recent studies have also explored detection, prevention, and resilience-oriented approaches for securing SV communication. Ustun et al. [8] proposed an artificial-intelligence-based intrusion detection system for IEC 61850 sampled values. More recently, Cibin et al. [9] proposed a hybrid statistical-deep-learning method for the detection, prevention, and source localization of SV injection attacks, whereas Mishchenko et al. [10] experimentally demonstrated that coordinated and physically consistent false-data injections against SV-based protection can remain stealthy and therefore advocated resilience measures such as trusted independent channels and cross-verification of SV data within protection logic. Related work has also explored broader data-driven anomaly-detection frameworks for IEC 61850 multicast traffic, including GOOSE and SV, using generative-AI and in-context-learning techniques [11], [12]. However, such methods are not 100% accurate and are best viewed as complementary to cryptographic protection as they are probabilistic rather than deterministic, depend on the representativeness of training or reference data, may incur false positives and false negatives, can degrade under unseen operating conditions or attack strategies, and generally do not provide direct per-packet data-

S.M.S. Hussain is with Electrical Engineering Department and Interdisciplinary Research Center for Sustainable Energy Systems (IRC-SES), King Fahd University of Petroleum & Minerals (KFUPM), Dhahran, Saudi Arabia. (email: muhammad.shaik@kfupm.edu.sa).

A. Ahmad and S.M. Farooq are with Vellore Institute of Technology (VIT), Vellore, India

M. Tayyab is with IRC-SES, KFUPM, Dhahran, Saudi Arabia.

origin authenticity and integrity guarantees.

IEC 62351-6 addresses security for fast IEC 61850 Layer-2 messages by recommending symmetric message-authentication mechanisms, most notably HMAC-SHA-256 and AES-GMAC, for GOOSE and SV traffic [13]. These standardized choices have motivated several implementation-oriented studies. For example, [14] proposed a fixed-latency architecture to secure GOOSE and sampled value messages in substation systems, while [4], [15] analyzed and implemented Message Authentication code (MAC) algorithms for SV message security and highlighted the practical tradeoffs among MAC algorithms in time-critical IEC 61850 communication. Nevertheless, the fundamental challenge remains that cryptographic protection must be applied to periodic multicast traffic that may be transmitted thousands of times per second. In [7], author proposed a lightweight optimized message authentication scheme for IEC 61850 sampled value messages in which only the time-varying fields are authenticated, whereas static fields are validated at the subscriber by comparison with previously received values. By reducing the authenticated input from the full SV payload to the dynamic portion of the stream, [7] lowers per-frame cryptographic cost while preserving the message-authentication objective. Even so, the computational burden remains challenging for higher-rate IEC 61869-9 profiles and for software-based or resource-constrained IEDs that may subscribe to multiple concurrent SV streams [3], [7].

For practical protocol deployment, the operational authentication algorithm should be lightweight and standard-aligned rather than ad hoc. Motivated by this requirement, this paper adopts Chaskey-12 [16] for IEC 61869-9 SV authentication and further proposes a compiler-optimized implementation to reduce execution latency.

All such MAC-based authentication schemes depend on a shared symmetric key. IEC 62351-9 addresses this requirement through a key-distribution-center (KDC)-based group-key-management architecture, and in this work the operational group-key-distribution logic is built around GDOI [17], [18]. The mutual authentication and authorization stage relies on the ISAKMP framework [19], while certificate-based trust establishment in currently deployed environments is typically realized using classical digital-signature mechanisms such as Rivest–Shamir–Adleman (RSA) and elliptic-curve schemes, and can be supported through standard transport-security tooling such as TLS [19], [20]. However, this classical asymmetric foundation is increasingly problematic for long-lived power-system assets. RSA relies on the computational hardness of integer factorization, whereas elliptic-curve digital signatures rely on the elliptic-curve discrete logarithm problem; both assumptions are vulnerable in the presence of quantum adversaries through Shor-type attacks [21]. More broadly, the power grid is now exposed to an emerging class of quantum-enabled cyber threats that may render currently deployed public-key mechanisms obsolete over the operational lifetime of utility infrastructure [22]–[24]. Recent IEC 61850 research has started to investigate post-quantum signatures for GOOSE communication, including Rainbow- and Falcon-based proposals [25], [26]. However, Rainbow has been seriously undermined by published attacks [27], while Falcon, although compact, still involves a comparatively complex signing process and implementation burden that make frequent per-message signing unattractive for high-sampling SV traffic [26]. These observations motivate a quantum-safe redesign of the initial trust-establishment stage for IEC 61850-based group-key distribution, while retaining lightweight symmetric authentication for operational SV frames. In this context, the present work adopts ML-KEM, now standardized by NIST in FIPS 203, to realize post-quantum-capable trust establishment for IEC 61869-9 SV communication [28].

Most prior cryptographic studies on SV have primarily emphasized feasibility in terms of computational delay, latency, or hardware/software implementation [7], [14]-[15], [4] whereas key establishment, runtime memory overhead is reported much less frequently, and rigorous security treatment of the overall scheme is seldom studied together with implementation results. Table I summarizes the positioning of the proposed work with respect to the most relevant state-of-the-art studies and standards.

Motivated by the above, this paper presents an integrated security framework for IEC 61869-9 SV communication that combines lightweight per-frame authentication for operational traffic with post-quantum-capable trust establishment for group-key distribution. The paper provides a security analysis of the proposed key-establishment procedure using a game-based treatment in the Quantum Random Oracle Model (QROM) and formal symbolic verification using AVISPA tool. It also presents an implementation of the complete security framework in C-language and evaluates its feasibility on a resource-constrained platform in terms of computational delay and memory utilization. Accordingly, the main contributions of this paper are fourfold.

- First, it develops a compiler-optimized Chaskey-12 implementation for low-latency authentication of IEC 61869-9 SV traffic.
- Second, it integrates ML-KEM into the GDOI-based trust-establishment stage for symmetric group-key provisioning.
- Third, it analyzes the security of the proposed key-establishment procedure in the QROM and verifies the protocol under the Dolev–Yao adversarial model.
- Fourth, a C-based publisher/subscriber prototype implements the proposed security framework on an experimental testbed and evaluates the computational and memory overhead across all eight IEC 61869-9 SV profiles.

The remainder of the paper is organized as follows. Section II introduces the background on IEC 61869-9 SV streams, standard message-authentication mechanisms, GDOI-based key management, and ML-KEM. Section III presents the proposed security framework. Section IV provides the security analysis and formal verification results. Section V reports the implementation details and performance evaluation. Section VI concludes the paper.

TABLE I
COMPARISON WITH STATE-OF-THE-ART SV SECURITY APPROACHES

| Ref | Main objective | Authentication | IEC 61869-9 SV support | Key establishment |
|---|---|---|---|---|
| [13] | Standardized protection for GOOSE/SV | ☑ | ☒ | ☒ |
| [14] | Fixed-latency secure GOOSE/SV architecture | ☑ | ☒ | ☒ |
| [4], [15] | Secure SV implementation and attack mitigation | ☑ | ☒ | ☒ |
| [7] | Lightweight optimized SV authentication | ☑ | ☒ | ☒ |
| [8], [9], [10], [11], [12] | Detection, prevention, or resilience against SV-related attacks | -- | ☒ | ☒ |
| This work | Real-time secure IEC 61869-9 SV communication | ☑ | ☑ | ☑ |

## II. BACKGROUND ON IEC 61869-9, IEC 62351-6, AND IEC-62351-9

This section summarizes the standards background and timing/security constraints that directly affect the proposed framework. First, the evolution from IEC 61850-9-2 and IEC 61850-9-2LE to IEC 61869-9 is discussed to clarify the SV publication context. Next, the MAC-based security mechanisms recommended in IEC 62351-6 are reviewed. Finally, the IEC 62351-9/GDOI group-key-management architecture is summarized to motivate the proposed ML-KEM-based trust-establishment stage.

### A. IEC 61869-9: SV Profile and Timing Analysis

IEC 61850-9-2 defines the communication mapping for sampled values over Ethernet within the IEC 61850 framework. It enables merging units to transmit digitized current and voltage measurements as multicast Layer-2 SV frames to multiple subscribing IEDs. However, early practical deployment required a more constrained and interoperable profile. The UCA International Users Group therefore introduced IEC 61850-9-2 Light Edition, commonly referred to as 9-2LE, as an implementation guideline that restricted the broader IEC 61850-9-2 mapping into a realizable process-bus profile for multi-vendor systems. The 9-2LE profile became widely used in early process-bus deployments because it defined practical conventions for synchronized current and voltage streaming from merging units to protection and control IEDs.

IEC 61869-9 subsequently standardized the digital interface for instrument transformers and consolidated SV publication behavior for modern digital substations. Compared with the broader communication-mapping role of IEC 61850-9-2 and the deployment-oriented role of 9-2LE, IEC 61869-9 explicitly defines the digital output measurement chain, merging function, synchronization requirements, dataset composition, and standardized SV publication profiles. Therefore, this paper treats IEC 61869-9 as the operative evaluation baseline, while IEC 61850-9-2 and 9-2LE are discussed as the communication and deployment lineage from which current process-bus SV practice evolved.

The timing burden is not determined by sampling rate alone. Packing multiple ASDUs in one frame reduces the frame rate and increases the interarrival interval, but it also increases the per-frame payload and, consequently, the amount of data that must be parsed and authenticated. Table II summarizes the eight IEC 61869-9 SV publication profiles considered in this paper.

The one-ASDU profiles impose the highest per-frame scheduling pressure because the subscriber has very limited time for reception, parsing, authentication, and forwarding to protection logic before the next frame arrives. The 96 kHz profile is the most demanding case, with an interarrival interval of only about 10.4 μs. In contrast, profiles with multiple ASDUs per frame provide a longer interarrival interval but increase the authenticated payload size. When an IED subscribes to multiple SV streams, frames from different streams compete for the same processing resources, further reducing the effective timing margin. For this reason, the proposed authentication scheme is evaluated across all eight IEC 61869-9 profiles rather than at a single representative operating point.

TABLE II
IEC 61869-9 STANDARDIZED SV PUBLICATION PROFILES AND INTERARRIVAL TIMES

| Sampling rate (Hz) | ASDUs / frame | Frames per second (fps) | Interarrival time |
|---|---|---|---|
| 4000 | 1 | 4000 | 250 μs |
| 4800 | 1 | 4800 | ~208 μs |
| 4800 | 2 | 2400 | ~416.7 μs |
| 5760 | 1 | 5760 | ~173.6 μs |
| 12800 | 8 | 1600 | 0.625 ms |
| 14400 | 6 | 2400 | ~416.7 μs |
| 15360 | 8 | 1920 | ~520.8 μs |
| 96000 | 1 | 96000 | ~10.4 μs |

### B. IEC 62351-6 MAC Security

IEC 62351 is the IEC Technical Committee 57 security series for power-system communications. IEC 62351-6 specifies cybersecurity mechanisms for IEC 61850 fast Layer-2 messages, including SV and GOOSE. The standard targets message modification, spoofing, replay, and man-in-the-middle manipulation by providing integrity and data-origin-authentication mechanisms for time-critical process-bus traffic [13]. The standard treats integrity and data-origin authenticity as core requirements and explicitly recognizes the timing constraints and device resource limitations of IED-class platforms. The earlier IEC 62351-6:2007 revision recommended RSA based digital signatures, but performance studies showed that RSA signature verification requires several milliseconds per operation-far exceeding both the 0.25 ms SV interarrival interval and the 3 ms IEC 61850 protection timing limit-making digital signatures operationally infeasible for high-rate process-bus security. The 2020 revision therefore adopted symmetric-key mechanisms specifically HMAC-SHA256 and AES-GMAC-as the recommended cryptographic primitives for protecting SV and GOOSE messages over a preshared group key.

For SV messages, IEC 62351-6:2020 specifies that the authentication value is calculated over the SV content beginning from the Ethernet EtherType field through the end of the SV ASDU. The resulting MAC value is appended to the message as a security extension, with additional fields supporting key rollover and security metadata. At the subscriber, the authentication value is recomputed over the same field range using the shared key, and the frame is discarded immediately on mismatch. This mechanism provides cryptographic integrity and data-origin authentication for every SV frame, but its per-packet overhead depends directly on the cost of applying the MAC over nearly the entire SV Protocol Data Unit (PDU).

A critical observation that enables a more efficient design is that SV streams are periodic and highly structured. In steady-state operation, only the sample counter (smpCnt) and the measurement payload (seqData) change from one frame to the next. Other fields-such as APPID, noASDU, svID, confRev, and synchronization indicators-remain constant throughout a valid SV stream. This structure allows a two-stage optimized authentication approach. First, the publisher computes the MAC only over smpCnt ‖ seqData and appends the resulting tag to the SV frame. The subscriber recomputes this MAC and discards the frame immediately on mismatch. Second, if the MAC verification succeeds, the subscriber validates the static fields by comparing them against cached expected values for that stream. This strategy preserves the security objective of IEC 62351-6 while eliminating repeated MAC computation over the large static portion of the SV PDU.

### C. IEC 62351-9 Key Management

IEC 62351-9 addresses the key-management requirement for secure IEC 61850 communication. Since HMAC and GMAC, are symmetric-key-based authentication mechanisms, publishers and subscribers must share a valid group key before protected SV communication can begin. This requirement is particularly important for SV because the protocol operates in multicast mode, where one publisher transmits measurement frames to multiple subscribing IEDs.

IEC 62351-9 introduces a Key Distribution Center based architecture for power-system equipment. In this work, the group-key-management process is built around the Group Domain of Interpretation mechanism. GDOI operates in two phases. In the first phase, each publisher and subscriber is mutually authenticated and authorized with the KDC, generation of a unique Long-Term Authentication Key (LTAK) for each participant. In the second phase, LTAK is used to securely distribute the operational group key (gk), which is subsequently used for MAC generation and verification over SV messages as shown in the Fig. 1 [18].

In legacy deployments, the initial authentication and authorization phase can rely on classical asymmetric cryptographic mechanisms through ISAKMP/TLS-oriented trust-establishment infrastructure. These mechanisms commonly use RSA or elliptic-curve-based public-key algorithms. However, RSA depends on the hardness of integer factorization, while elliptic-curve schemes depend on the elliptic-curve discrete logarithm problem. Both assumptions are vulnerable to sufficiently powerful quantum adversaries. This is a critical concern for digital-substation assets because such infrastructure has a long operational lifetime.

Recent work has explored post-quantum digital signatures for IEC 61850 communication. However, signature-based approaches are not attractive for frequent per-frame protection of high-rate SV traffic because they impose significant computational and communication overhead. Therefore, this paper does not attempt to sign every SV frame. Instead, it retains lightweight symmetric authentication for operational SV messages and introduces ML-KEM into the GDOI trust-establishment stage to provide quantum-safe key establishment before group-key distribution.

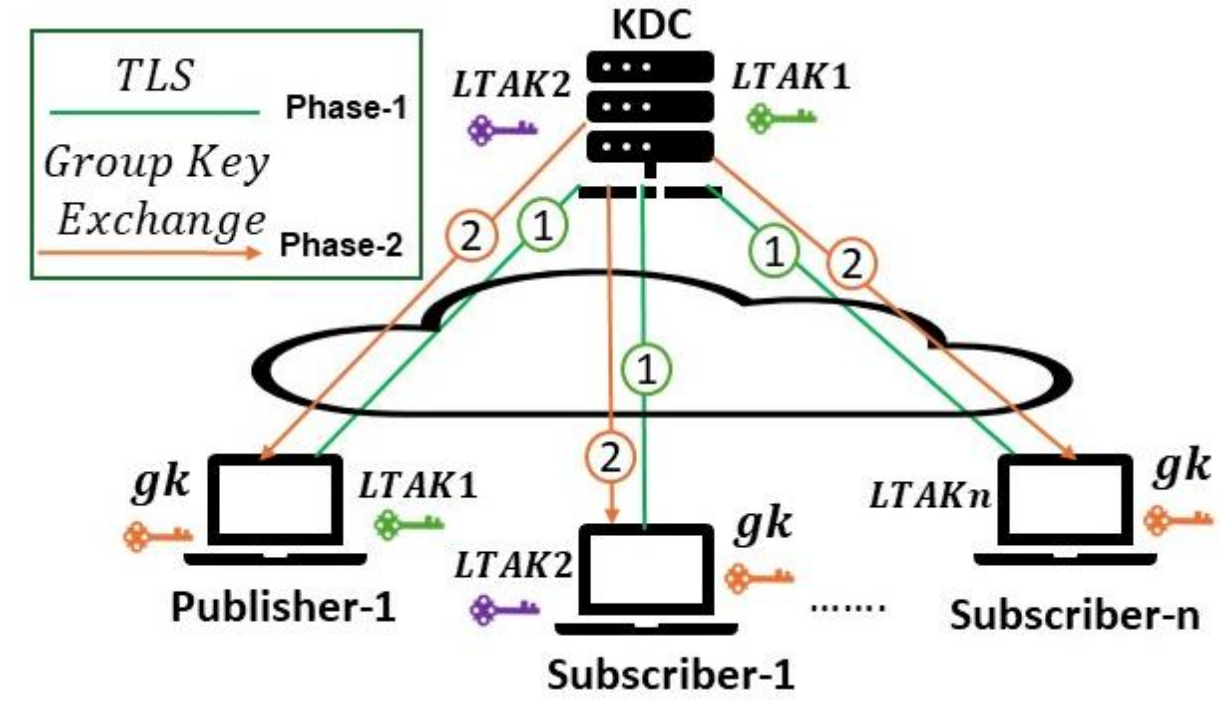

**Fig. 1.** GDOI based authentication and key distribution.

## III. Proposed Security Scheme

This section presents the proposed integrated security framework. The first component is an ML-KEM-based two-party authentication and key-establishment protocol that replaces the classical trust-establishment stage between the KDC and each authorized IEC 61850 participant. The second component is a Chaskey-12-based, field-selective MAC scheme for operational SV traffic. Together, these mechanisms address the two complementary layers of secure SV communication: secure key establishment and low-latency per-frame authentication. To the best of the authors' knowledge, this is among the first implementation-oriented studies to jointly evaluate lightweight Chaskey-12 authentication and ML-KEM-based post-quantum key establishment for IEC 61869-9 SV communication.

### A. Proposed ML-KEM based two party authentication protocol

The proposed protocol uses the NIST-standardized ML-KEM Mechanism for authentication of producers/subscribers with KDC. The quantum safe ML-KEM replaces TLS mechanism recommended by GDOI mechanism in its first phase to authenticate the communicating entities that ensures sharing of LTAKs for each participant. Further, LTAKs are exploited to share a group key (*gk*) in the second phase. The group key (*gk*) is further used in the generation of Message

Authentication Code (MAC) based on Chaskey-12 cryptographic lightweight algorithm to achieve the message integrity of SV messages. Fig. 2 shows the illustration of ML-KEM authentication at first phase of GDOI and secure communication of SV messages using Chaskey-12.

A two-party post-quantum authentication and key establishment protocol between a trusted KDC and publisher/Subscriber IEDs, designed using the NIST-standardized ML-KEM mechanism as specified in FIPS 203 [28]. The proposed system consists of two participating entities: the KDC, which functions as a trusted authority responsible for authentication and session key establishment, and the publisher/subscriber. The authentication and key establishment procedure is executed over a public and insecure communication channel, while secure channels are assumed only during the offline registration phase. The adversarial capabilities are modelled according to the Dolev–Yao (DY) threat model, wherein the adversary has full control over the public channel but cannot compromise cryptographic primitives.

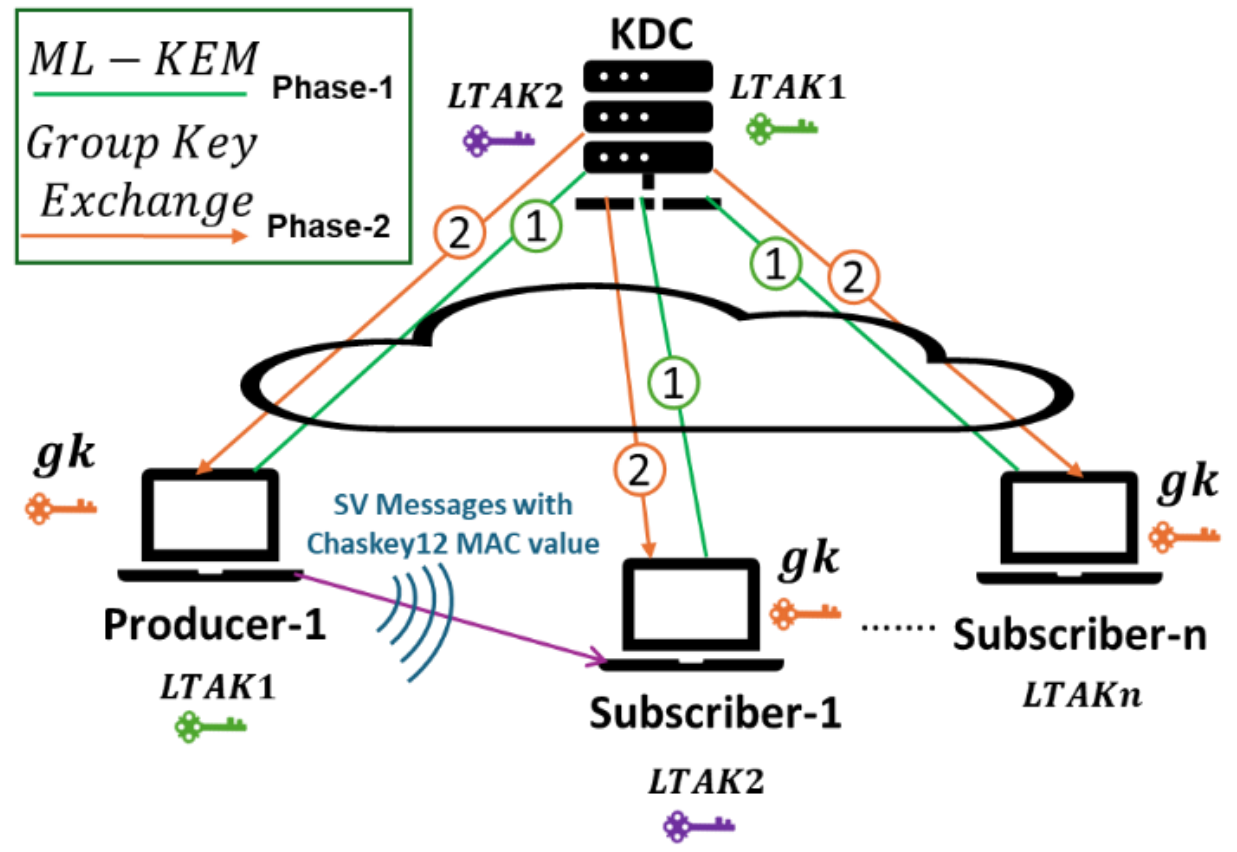

**Fig. 2.** GDOI based ML-KEM authentication and secure communication using chaskey-12.

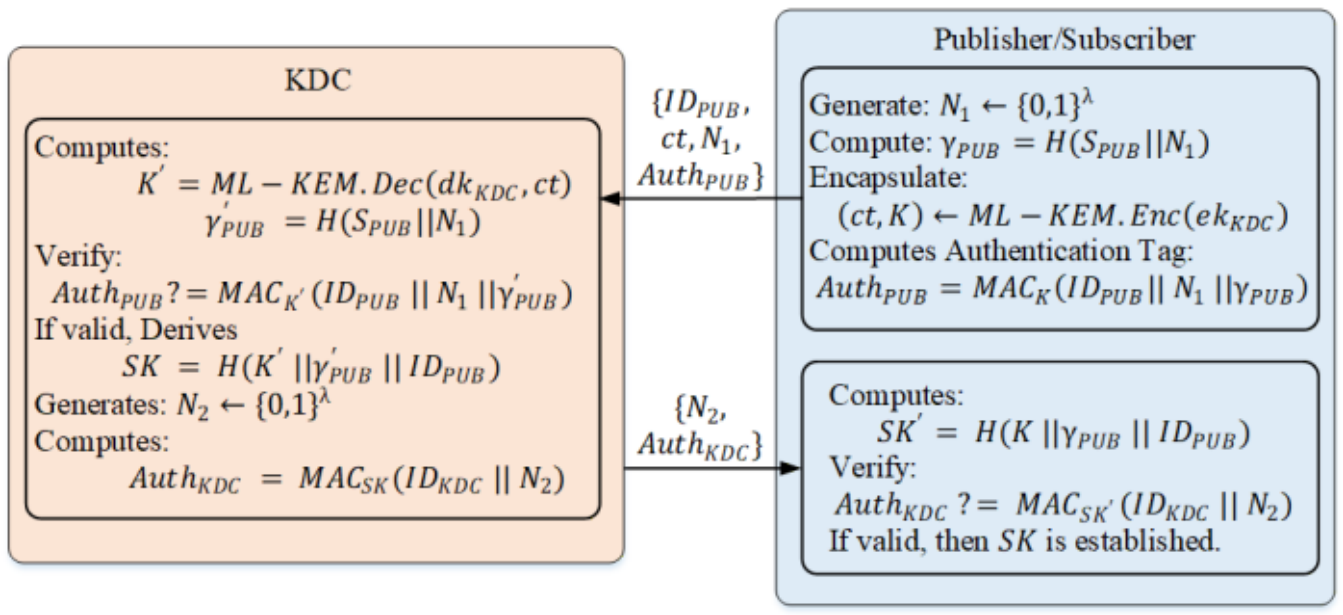

**Fig. 3.** ML-KEM-based two-party authentication protocol.

The ML-KEM-based two-party authentication protocol is shown in Fig. 3. The protocol employs ML-KEM.KeyGen(), ML-KEM.Enc(), and ML-KEM.Dec() for post-quantum key encapsulation, a secure cryptographic hash function H(·), and a message authentication code MAC_K(·), with the concatenation operator denoted by ||. The proposed protocol is structured into three distinct phases, namely the system setup phase, the registration phase, and the authentication and key establishment phase, each of which is described in detail in the subsequent sections.

1) **System Setup Phase**
The KDC executes the ML-KEM key generation algorithm to obtain a public–private key pair $(ek_{KDC}, dk_{KDC})$. The public key $ek_{KDC}$ is published system-wide, while the private key $dk_{KDC}$ is securely stored by the KDC. The KDC also generates and securely stores a high-entropy master secret, referred to as $KDC_{master}$, which is used exclusively during device registration.

2) **Registration Phase**
The Publisher/Subscriber IED submits its identity $ID_{PUB}$ to the KDC through a secure offline channel. Upon successful verification, the KDC derives a unique long-term secret $S_{PUB} = H(ID_{PUB} || KDC_{master})$ for the Publisher/Subscriber IED. The derived secret is securely transferred to the Publisher/Subscriber IED and stored locally. $S_{PUB}$ is never transmitted during online authentication.

3) **Authentication and Key Establishment**
This phase is executed over a public channel.
Step1: The Publisher/Subscriber IED generates a fresh nonce $N_1$ to ensure session freshness. Using its long-term secret and the nonce, the Publisher/Subscriber IED computes a credential-dependent masking value $\gamma_{PUB} = H(S_{PUB}||N_1)$. The Publisher/Subscriber IED then executes the ML-KEM encapsulation algorithm using the KDC's public key, producing a ciphertext $ct$ and a shared secret $K$. An authentication tag $Auth_{PUB}$ is computed over the identity, nonce, and masking value using the shared secret as the MAC key. The Publisher/Subscriber IED transmits $\{ID_{PUB}, ct, N_1, Auth_{PUB}\}$to the KDC.
Step 2: Upon receiving the request, the KDC decapsulates the ciphertext using $dk_{KDC}$ to recover $K'$. The KDC independently recomputes the masking value $\gamma'_{PUB}$ using $S_{PUB}$ and $N_1$. The KDC verifies the authenticity of the Publisher/Subscriber IED by validating the received authentication tag $Auth_{PUB} ?= MAC_{K'}(ID_{PUB} || N_1 || \gamma'_{PUB})$. If verification succeeds, the KDC derives a session key $SK = H(K' || \gamma'_{PUB} || ID_{PUB})$.
Step 3: The KDC generates a fresh nonce $N_2$ for key confirmation. Using the derived session key $SK$, the KDC computes a response authentication tag $Auth_{KDC} = MAC_{SK}(ID_{KDC} || N_2)$. The KDC sends $\{N_2, Auth_{KDC}\}$ to the Publisher/Subscriber IED.
Step 4: Upon receipt, the Publisher/Subscriber IED independently derives the session key $SK' = H(K || \gamma_{PUB} || ID_{PUB})$. The Publisher/Subscriber IED verifies the authentication tag using the derived session key $Auth_{KDC} ?= MAC_{SK'}(ID_{KDC} || N_2)$. Successful verification confirms both mutual authentication and correct session key establishment.

### *B. Chaskey-12 Authentication*

The authentication mechanism for operational IEC 61869-9 SV traffic must verify every received frame while satisfying the strict interarrival constraints of the selected SV profile. In this work, the selective-field MAC-generation strategy follows [7] and is therefore adopted rather than claimed as a new contribution. Specifically, only the time-varying portion of the SV message is authenticated cryptographically, while the static stream fields are verified through cached comparison after successful MAC verification. Let M = smpCnt || seqData denote the dynamic authentication input associated with an SV frame. The remaining invariant stream descriptors, such as APPID, svID, noASDU, confRev, and synchronization-related indicators, are treated as cached fields and are validated separately at the subscriber side [7].

Let (gk) denote the symmetric group key distributed through the GDOI procedure. For each outgoing SV frame, the publisher computes a 128-bit authentication tag $T = Chaskey - 12_{gk}(M)$ and appends $T$ to the frame. Upon reception, the subscriber recomputes the tag over the received dynamic fields using the same group key. If the recomputed tag does not match the received tag, the frame is discarded immediately. Only when the tag verification succeeds are the cached static fields compared against the expected stream template. This two-stage verification preserves the integrity and data-origin-authentication objective while avoiding repeated MAC computation over the invariant part of the SV protocol data unit [7].

The proposed contribution of this paper is not the selective-field framing itself, but the use of Chaskey-12 as the lightweight operational MAC for IEC 61869-9 SV traffic together with a compiler-optimized software realization. Chaskey-12 is attractive for this setting because it is based on simple Addition–Rotation–XOR (ARX) operations over 32-bit words and is therefore well suited to embedded and resource-constrained execution platforms [16].

To further reduce runtime latency, this paper proposes a compiler-optimized implementation of Chaskey-12. The optimization does not modify the cryptographic algorithm, its round structure, the key schedule, or the tag-generation procedure. Instead, performance is improved entirely at compilation time. Loop unrolling is applied to repetitive permutation routines in order to reduce loop-control overhead and expose more independent operations. Function inlining is used for short helper routines to eliminate call/return overhead and to enable additional cross-boundary compiler optimizations. Instruction scheduling is exploited to improve instruction-level parallelism and reduce pipeline stalls on the target processor. Finally, link-time optimization is used to enable whole-program analysis, cross-module inlining, and elimination of unnecessary code, thereby producing a more efficient binary implementation.

These compiler-driven optimizations reduce control-flow overhead, minimize function-call latency, and improve processor utilization without changing the cryptographic functionality of Chaskey-12. As a result, the optimized implementation preserves portability and security equivalence with the baseline Chaskey-12 algorithm while providing lower computational delay for real-time SV authentication. Section V evaluates this gain experimentally across the standardized IEC 61869-9 SV packet profiles.

## IV. Security Analysis of Proposed Scheme

The proposed scheme requires security validation from both analytical and formal perspectives. The integration of an ML-KEM-based authentication and key-establishment procedure with lightweight message authentication for time-critical SV traffic calls for a security treatment that addresses post-quantum resilience, session-key secrecy, mutual authentication, and resistance to common network attacks. This section therefore first analyzes the protocol in the QROM and then presents AVISPA-based formal verification under the Dolev-Yao model.

### *A. Security Analysis using QROM*

The QROM extends the classical Random Oracle Model to the post-quantum setting by allowing adversaries to issue quantum superposition queries to hash functions modelled as random oracles. In the QROM, a hash function $H: \{0,1\}^* \rightarrow \{0,1\}^l$ is treated as an idealized random function that can be queried both classically and in quantum superposition by a quantum polynomial-time (QPT) adversary. This model captures the realistic capability of quantum adversaries to evaluate cryptographic hash functions on multiple inputs simultaneously using quantum queries [29].

Security proofs in the QROM require careful handling of oracle programming, as the challenger may reprogram the random oracle at points that are computationally hidden from the adversary, while ensuring that such reprogramming remains indistinguishable except with negligible probability. Established techniques show that, provided the programmed inputs are not queried by the adversary prior to reprogramming, the adversary's view remains statistically or computationally unchanged.

The QROM has become the standard framework for analyzing post-quantum cryptographic protocols that rely on hash functions, including lattice-based constructions such as ML-KEM, whose security reductions are proven under the hardness of the Module Learning With Errors (MLWE) problem. Consequently, all hash functions used in the proposed protocol are modelled as quantum random oracles, and the adversary is assumed to be a QPT adversary with full access to these oracles.

1) ***Theorem***

Assuming that the Module Learning With Errors (MLWE) problem is computationally hard for any quantum polynomial-time (QPT) adversary, the ML-KEM scheme is IND-CCA secure in the QROM, and the employed message authentication code (MAC) is existentially unforgeable under chosen-message attacks, the proposed two-party authentication protocol achieves authenticated key establishment (AKE) with session key indistinguishability against any QPT adversary in the QROM.

2) ***Proof***

The proof proceeds via a sequence of games. Let $\mathcal{A}$ be a quantum polynomial-time adversary participating in the authenticated key establishment experiment, whose goal is to distinguish the real session key established between the KDC and the Publisher/Subscriber IED from a uniformly random key.

**a) Game₀ (Real Protocol Execution)**

This game corresponds to a real execution of the proposed protocol. The session key is computed as

$$SK = H(K||\gamma_{PUB}||ID_{PUB})$$

where $K$ is the shared secret obtained from ML-KEM encapsulation and

$$\gamma_{PUB} = H(S_{PUB}||N_1)$$

with $S_{PUB}$ denoting the long-term secret of the Publisher/Subscriber IED and $N_1$ a fresh nonce. All hash functions are modelled as quantum random oracles.

Let $\mathrm{Adv}_{\mathcal{A}}^{Game_0}$ denote the adversary's advantage in this game.

**b) Game₁ (Authentication Forgery Elimination)**

In this game, the challenger aborts if the adversary forges a valid authentication tag without knowledge of the corresponding secret key. Since authentication tags are generated using a secure MAC keyed with the ML-KEM shared secret, any successful forgery implies a violation of MAC existential unforgeability. Therefore, the difference between Game₀ and Game₁ is negligible.

$$|P_r[Game_0] - P_r[Game_1]| \leq negl(\lambda)$$

**c) Game₂ (Replacement of the ML-KEM Shared Secret):**

Here, the ML-KEM-derived shared secret KKK is replaced with a uniformly random string of the same length. By the IND-CCA security of ML-KEM under the MLWE assumption, any QPT adversary capable of distinguishing Game₁ from Game₂ can be used to break the IND-CCA security of ML-KEM in the QROM. Hence,

$$|P_r[Game_1] - P_r[Game_2]| \leq \mathrm{Adv}_{\mathcal{A}}^{ML-KEM}$$

**d) Game₃ (Random Oracle Programming):**

In this game, the quantum random oracle is programmed so that the session key is derived from uniformly random inputs. Since the long-term secret $S_{PUB}$ and nonce $N_1$ remain unknown to the adversary, the value $\gamma_{PUB} = H(S_{PUB}||N_1)$ is computationally indistinguishable from random. Standard QROM programming arguments ensure that the adversary's view remains unchanged except with negligible probability:

$$|P_r[Game_2] - P_r[Game_3]| \leq negl(\lambda)$$

**e) Game₄ (Key Indistinguishability)**

In the final game, the session key is replaced with a uniformly random value independent of the protocol execution. Consequently, the adversary's view is statistically independent of the challenge bit, and Pr[b=b′]= $\frac{1}{2}$. By applying the triangle inequality over the sequence of games, the overall advantage of the adversary is bounded as

$$\mathrm{Adv}_{\mathcal{A}}^{AKE} \leq \mathrm{Adv}_{\mathcal{A}}^{ML-KEM} + \mathrm{negl}(\lambda).$$

### *B. Security Analysis Using AVISPA tool*

The proposed protocol was formally verified using the AVISPA tool under the Dolev–Yao threat model [30]. The protocol was specified in HLPSL and analyzed using both the On-the-Fly Model Checker (OFMC) and the Automated Tree Search Engine (ATSE) back ends. In both cases, the analysis returned a "SAFE" result, confirming that the protocol satisfies the specified secrecy and authentication goals and is resistant to replay and man-in-the-middle attacks within the considered adversarial model. The verification outcomes produced by the OFMC and ATSE back ends are illustrated in Fig. 4 and Fig. 5, respectively, which demonstrate the absence of feasible attack traces under the defined security goals.

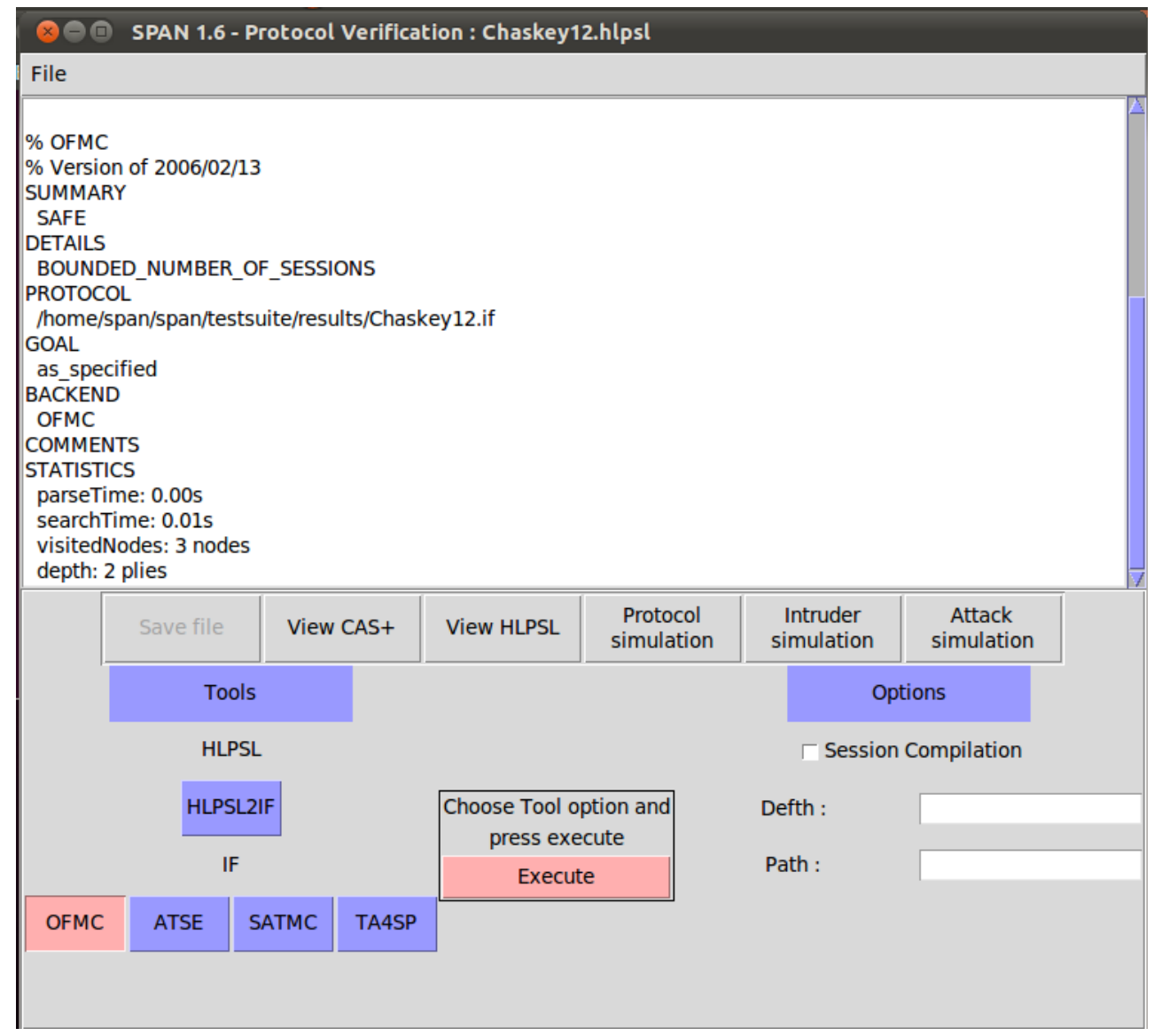

**Fig. 4.** AVISPA verification result for the proposed protocol using the OFMC backend.

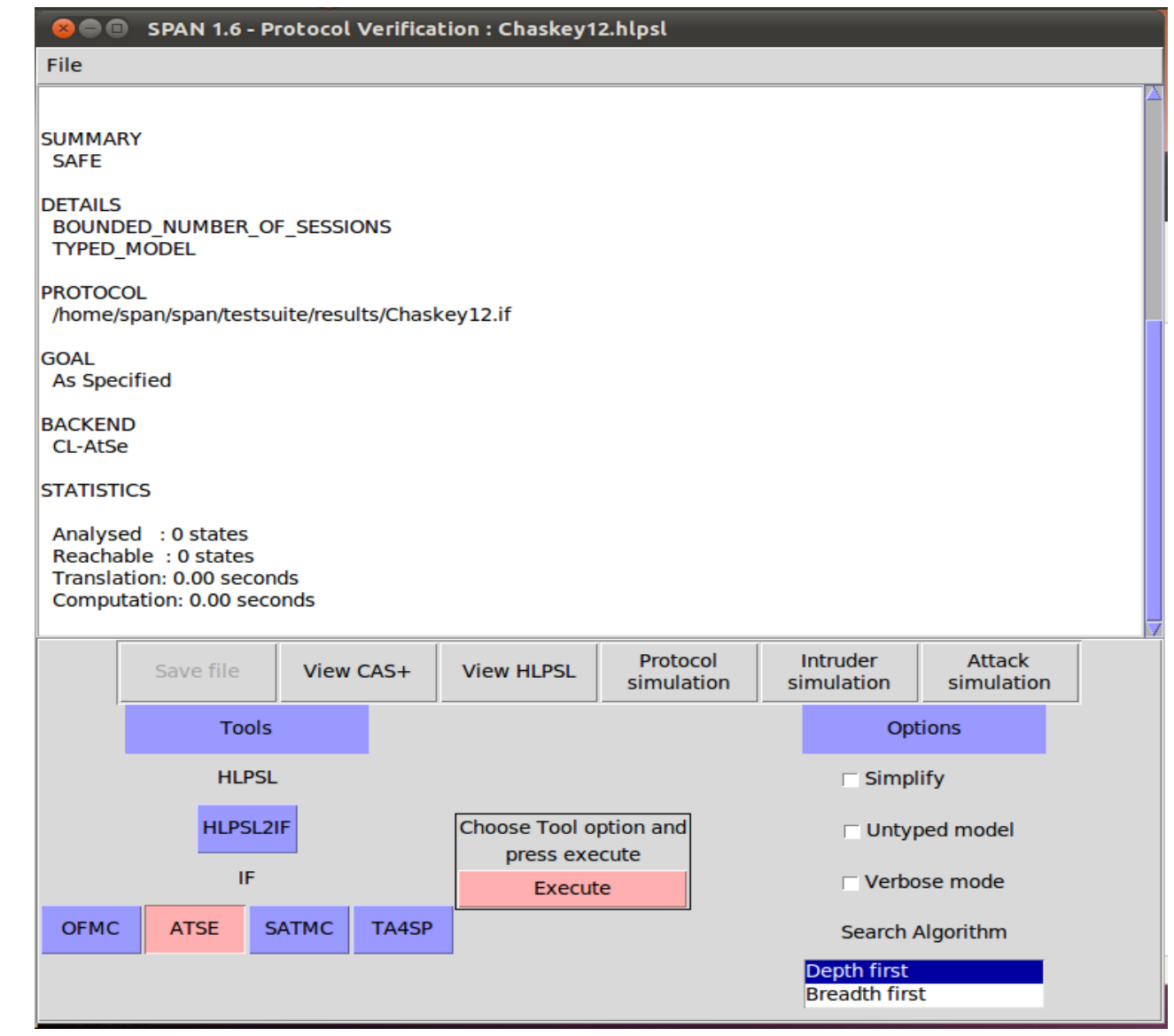

**Fig. 5**. AVISPA verification result for the proposed protocol using the ATSE backend.

## V. EXPERIMENTAL IMPLEMENTATION AND PERFORMANCE EVALUATION

This section evaluates the proposed security framework from both algorithm-level and implementation-level perspectives across the IEC 61869-9 SV packet profiles summarized in Table II.

### A. Implementation and Benchmarking Setup

A C-based prototype library was developed to publish and subscribe to secure SV messages using the proposed authentication scheme. Following the laboratory-validation approach commonly adopted in secure SV implementation studies [4], [7], [15], the prototype was realized on two Raspberry Pi nodes interconnected through a Layer-2 Ethernet switch and a terminal PC (laptop), as shown in Fig. 6. The two nodes emulated two distinct IEDs, namely an SV publisher representing the merging-unit side and an SV subscriber representing the protection-and-control IED side. The terminal PC emulates the KDC. The publisher generates IEC 61869-9 SV frames and appends a 16-byte Chaskey-12 authentication value, whereas the subscriber receives the frames, recomputes the authentication value, and accepts the message only after successful verification. Wireshark is used to capture the exchanged traffic and verify the on-wire structure of the security-protected SV frame. Fig. 7 shows a representative Wireshark capture of a 4800-Hz, 2-ASDU IEC 61869-9 SV message. The capture confirms that the standard SV fields are preserved and that the secure message is extended by appending the Chaskey-12 MAC value in the security field. A comparatively modest hardware platform was selected intentionally so that the feasibility of the proposed scheme could be assessed on resource-constrained endpoints; if the timing budget is satisfied on this platform, deployment on contemporary IED-class hardware with higher processing capability becomes less demanding.

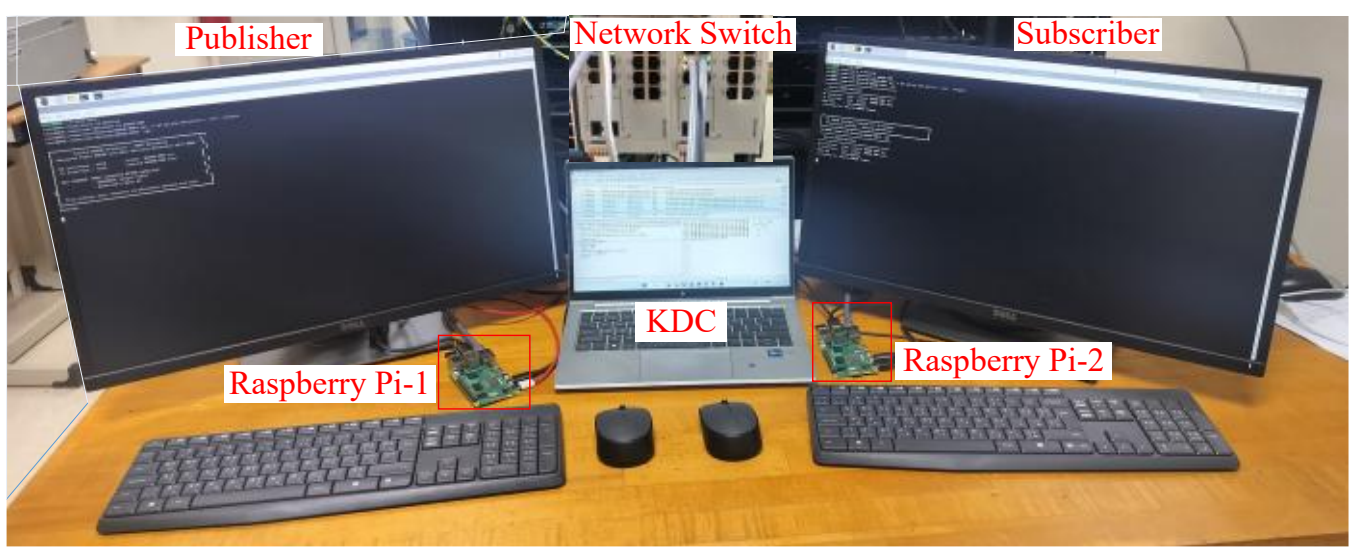

**Fig. 6.** Experimental testbed for Secure IEC 61869-9 SV communication.

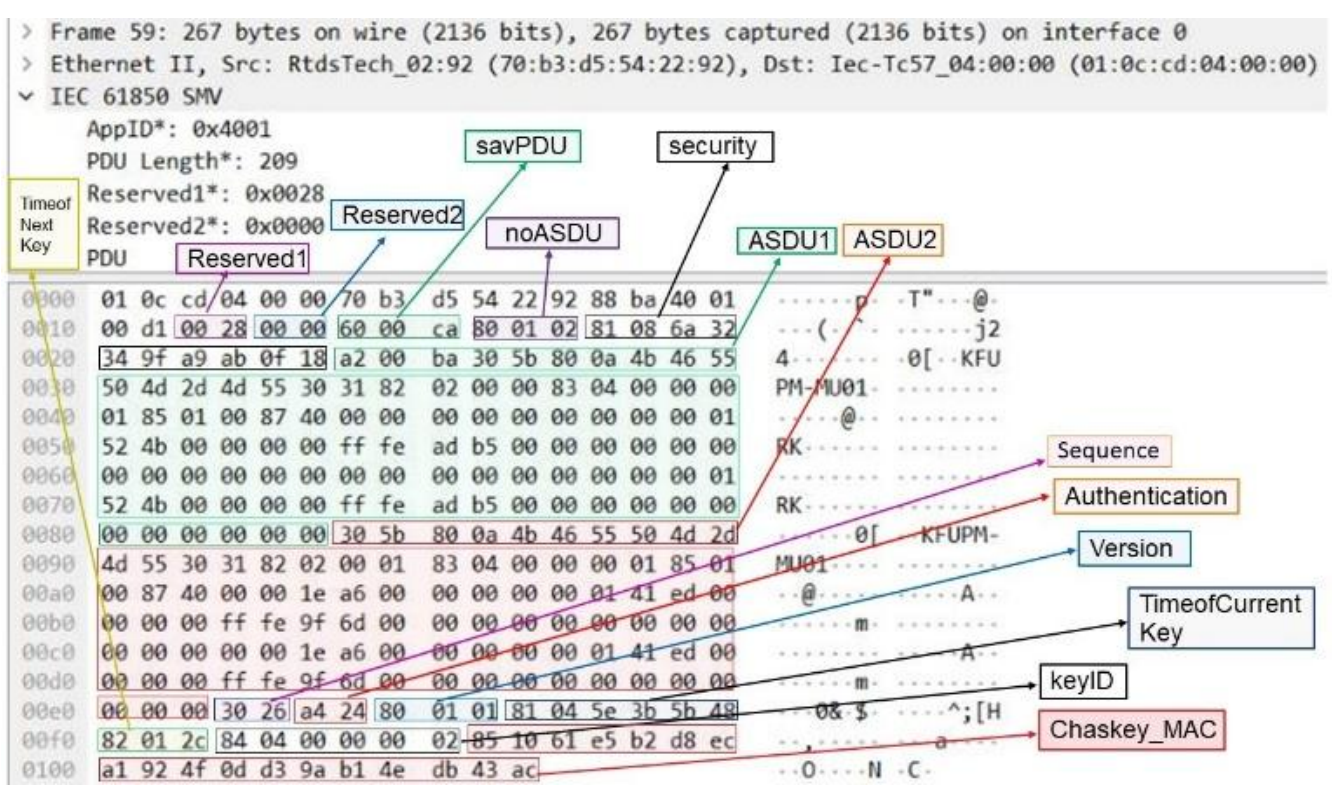

**Fig. 7.** Wireshark capture of a Secure IEC 61869-9 SV frame.

In addition to the lightweight authentication framework, the post-quantum key-establishment phase is also implemented. The ML-KEM implementation is realized using the liboqs library [31], which provides standardized implementations of NIST post-quantum cryptographic algorithms. The liboqs API was integrated into the authentication module to perform ML-KEM key generation, encapsulation, and decapsulation during the first phase of the proposed GDOI-based authentication procedure.

The timing and memory benchmark of the MAC computation was conducted on a Raspberry Pi running a PREEMPT_RT real-time kernel with OpenSSL 3.5.0. To reduce operating-system-induced timing variation, the process address space was locked using mlockall() system call, the benchmark thread executed under SCHED_FIFO real-time scheduling policy with priority 80, and an OpenSSL warm-up call was issued before timing so that library initialization and buffer allocation occurred outside the measured interval. These measures reduced scheduler-induced jitter and improved repeatability of the latency measurements. For each SV profile and MAC implementation, the computational timing experiment was repeated 100,000 times, and the arithmetic average is reported in Table III. The reported average wall-clock time denotes the real elapsed time required to complete the MAC computation, whereas VmRSS denotes the resident set size, i.e., the portion of process memory physically resident in RAM during execution.

### B. Comparative Computational-Time and Memory Results

For the 4000-Hz 1-ASDU, 4800-Hz 1-ASDU, 5760-Hz 1-ASDU, and 96000-Hz 1-ASDU profiles, the MAC input size is identical at 120 bytes. As a result, the measured execution time of each algorithm is the same across these four cases, even though the system-level timing pressure differs because the interarrival interval changes with the publication rate. HMAC-256 requires 0.009784 ms with a VmRSS of 9280 KB, whereas AES-GMAC-256 reduces the execution time to 0.004925 ms at a slightly higher VmRSS of 9416 KB. Blake-2s, Chaskey-12, and compiler-optimized Chaskey-12 reduce the execution time further to 0.002700, 0.000565, and 0.000525 ms, respectively, while maintaining VmRSS values in the narrow range of 2200-2204 KB. These results show that the lightweight implementations provide both lower latency and significantly smaller resident memory footprints than the HMAC-256 and AES-GMAC-256 baselines.

For the multi-ASDU profiles, the MAC input size increases from 215 bytes to 775 bytes, and all algorithms exhibit higher execution time. Chaskey-12 and compiler-optimized Chaskey-12 remain the fastest alternatives, reaching only 0.002821 ms and 0.002113 ms, respectively, for the 775-byte profiles. Thus, for the largest SV packet size, compiler-optimized Chaskey-12 is approximately 5.9 times faster than HMAC-256 and 4.5 times faster than AES-GMAC-256.

The memory results are also consistent across the evaluated packet sizes. HMAC-256 and AES-GMAC-256 maintain VmRSS values of 9280 KB and 9416 KB, respectively, whereas Blake-2s and both Chaskey-12 implementations remain close to 2200-2204 KB. This indicates that the resident memory footprint is primarily determined by the implementation and library dependencies rather than by the SV payload size over the measured range.

TABLE III
MESSAGE AUTHENTICATION TIME FOR IEC 61869-9 SV PACKET PROFILES

| Packet | MAC Input (bytes) | HMAC-256 (32B MAC) | | AES-GMAC-256 (32B MAC) | | Blake-2s (32B MAC) | | Chaskey-12 (16B MAC) | | Chaskey-12 (16B MAC) Compiler optimization | |
|---|---|---|---|---|---|---|---|---|---|---|---|
| | | VmRSS (KB) | Avg wall-clock time (ms) | VmRSS (KB) | Avg wall-clock time (ms) | VmRSS (KB) | Avg wall-clock time (ms) | VmRSS (KB) | Avg wall-clock time(ms) | VmRSS (KB) | Avg wall-clock time (ms) |
| 4000-1ASDU | 120 | 9280 | 0.009784 | 9416 | 0.004925 | 2204 | 0.002700 | 2200 | 0.000565 | 2204 | 0.000525 |
| 4800-1ASDU | 120 | 9280 | 0.009784 | 9416 | 0.004925 | 2204 | 0.002700 | 2200 | 0.000565 | 2204 | 0.000525 |
| 5760-1ASDU | 120 | 9280 | 0.009784 | 9416 | 0.004925 | 2204 | 0.002700 | 2200 | 0.000565 | 2204 | 0.000525 |
| 96000-1ASDU | 120 | 9280 | 0.009784 | 9416 | 0.004925 | 2204 | 0.002700 | 2200 | 0.000565 | 2204 | 0.000525 |
| 4800-2ASDU | 215 | 9280 | 0.010031 | 9416 | 0.00555 | 2204 | 0.003789 | 2200 | 0.000873 | 2204 | 0.000673 |
| 14400-6ASDU | 589 | 9280 | 0.011623 | 9416 | 0.008192 | 2204 | 0.007647 | 2200 | 0.002165 | 2204 | 0.001619 |
| 12800-8ASDU | 775 | 9280 | 0.012417 | 9416 | 0.009506 | 2204 | 0.009627 | 2200 | 0.002821 | 2204 | 0.002113 |
| 15360-8ASDU | 775 | 9280 | 0.012417 | 9416 | 0.009506 | 2204 | 0.009627 | 2200 | 0.002821 | 2204 | 0.002113 |

Therefore, the lightweight standalone implementations provide a substantially smaller runtime footprint than the OpenSSL-based HMAC-256 and AES-GMAC-256 baselines.

Overall, Table III shows that the compiler-optimized Chaskey-12 implementation provides the best latency-memory tradeoff among the evaluated methods. AES-GMAC-256 remains relevant when standards compliance or hardware-assisted acceleration is prioritized, while HMAC-256 provides broad compatibility at the cost of the highest computational delay and memory footprint. The key conclusion is therefore implementation-aware: Chaskey-12 is attractive not only because of its lightweight design, but also because compiler-level optimization further reduces its execution latency without modifying the cryptographic algorithm.

To evaluate the practicality of the proposed post-quantum authentication framework, the computational overhead of the ML-KEM operations was measured using the liboqs implementation on the experimental platform. The benchmark considered the three NIST-standardized parameter sets, namely ML-KEM-512, ML-KEM-768, and ML-KEM-1024. The average execution times for key generation, encapsulation, and decapsulation are summarized in Table IV.

TABLE IV
IEC 61869-9 SV TOTAL PROCESSING TIME PER MESSAGE

| Algorithm | KeyGen (µs) | Encapsulation (µs) | Decapsulation (µs) |
|---|---|---|---|
| ML-KEM-512 | 43.624 | 48.394 | 48.466 |
| ML-KEM-768 | 67.630 | 72.043 | 76.316 |
| ML-KEM-1024 | 97.556 | 104.556 | 113.490 |

### C. *Publisher-Side Processing Delay Against IEC 61869-9 Interarrival Time*

Table V evaluates the complete publisher-side processing delay for each IEC 61869-9 SV profile. The total delay is calculated from the instant at which the digital measurement values are available until the complete security-protected SV message is delivered to the Ethernet port. Table V shows the practical processing budget associated with SV frame construction, authentication, security-field insertion, and delivery to the network interface.

The results show that the proposed implementation satisfies the interarrival-time requirement for seven of the eight standardized profiles. For the 4000-Hz, 4800-Hz, and 5760-Hz 1-ASDU profiles, the total processing delay is 71.93 µs, which remains below the corresponding interarrival times of 250 µs, ~208 µs, and ~173.6 µs, respectively. For the 4800-Hz 2-ASDU, 14400-Hz 6-ASDU, and 12800-Hz / 15360-Hz 8-ASDU, profiles, the total delays of 84.27 µs, 126.01 µs, and 151.79 µs also remain within the available interarrival-time budgets. Hence, the proposed secure SV processing path is feasible for these standardized profiles on the evaluated resource-constrained platform.

The only exception is the 96000-Hz, 1-ASDU profile. Although its MAC input size is only 120 bytes, its interarrival interval is approximately 10.4 µs, which is much smaller than the measured total processing delay of 71.93 µs. This result indicates that the 96-kHz SV profile requires additional implementation support, such as hardware-assisted authentication, more aggressive real-time optimization, parallel processing, or dedicated IED-class hardware.

TABLE V
IEC 61869-9 SV TOTAL PROCESSING TIME PER MESSAGE

| Sampling rate (Hz) | ASDUs / frame | Interarrival time | Total delay (µs) | Applicability |
|---|---|---|---|---|
| 4000 | 1 | 250 µs | 71.93 | ☑ |
| 4800 | 1 | ~208 µs | 71.93 | ☑ |
| 4800 | 2 | ~416.7 µs | 84.27 | ☑ |
| 5760 | 1 | ~173.6 µs | 71.93 | ☑ |
| 12800 | 8 | 0.625 ms | 151.79 | ☑ |
| 14400 | 6 | ~416.7 µs | 126.01 | ☑ |
| 15360 | 8 | ~520.8 µs | 151.79 | ☑ |
| 96000 | 1 | ~10.4 µs | 71.93 | ☒ |

## VI. CONCLUSION

This paper presented an integrated security framework for IEC 61869-9 SV communication that combines lightweight message authentication with post-quantum-capable key establishment. The paper first analyzed the eight standardized IEC 61869-9 SV profiles and showed that their different ASDU packing structures, packet sizes, frame rates, and interarrival times create highly diverse real-time constraints for subscriber IEDs. To address this challenge without overloading the process-bus timing budget, the paper adopted field-selective authentication, in which only the time-varying SV fields are authenticated cryptographically and static fields are validated through cached comparison. Chaskey-12 was adopted as the MAC primitive due to its lightweight ARX design and suitability for embedded platforms, and an ML-KEM-based two-party authentication and key-establishment protocol was introduced to strengthen the key-management layer against quantum-era threats. The protocol was analyzed in the QROM and formally verified using AVISPA, where the OFMC and ATSE back ends confirmed the specified secrecy and authentication goals.

The experimental results confirm the feasibility of the proposed framework on a resource-constrained process-bus platform. A C-based publisher/subscriber prototype was implemented using two Raspberry Pi nodes, and Wireshark packet inspection verified that the Chaskey-12 MAC value is appended to the security-protected IEC 61869-9 SV frame. The MAC computation results, averaged over 100,000 executions, show that compiler-optimized Chaskey-12 achieves the lowest authentication time across all evaluated packet profiles while maintaining a compact memory footprint. The publisher-side total processing delay, calculated from the instant at which digital values are available until the complete secure SV frame is delivered to the Ethernet port, also satisfies the interarrival-time constraints for all evaluated profiles except the 96-kHz, 1-ASDU case. Therefore, the proposed framework provides a practical and scalable security solution for most standardized IEC 61869-9 SV operating profiles, while the 96-kHz profile may require hardware acceleration, tighter real-time scheduling, or dedicated communication support for deployment under the same security processing model.

## References

[1] L. Yang, P. A. Crossley, A. Wen, R. Chatfield, and J. Wright, “Design and performance testing of a multivendor IEC61850-9-2 process bus based protection scheme,” IEEE Trans. Smart Grid, vol. 5, no. 3, pp. 1159–1164, 2014, doi: 10.1109/TSG.2013.2277940

[2] C. Brunner, Lang Gerhard, L. Frederic, and S. Fred, “Implementation guideline for digital interface to instrument transformers using IEC 61850-9-2,” UCA Int. Users Gr., vol. 1, no. Figure 4, p. 31, 2004.

[3] Instrument transformers-Part 9: Digital interface for instrument transformers. IEC 61869-9:2021.

[4] S. M. S. Hussain, M. A. Aftab, S. M. Farooq, I. Ali, T. S. Ustun, and C. Konstantinou, “An Effective Security Scheme for Attacks on Sample Value Messages in IEC 61850 Automated Substations,” IEEE Open Access J. Power Energy, vol. 10, no. 2, pp. 304–315, 2023, doi: 10.1109/OAJPE.2023.3255790.

[5] S.P. Singh, A. Sharma, S. Jayatilleke, S. Battula, D. Alahakoon, “Securing sustainable digital substations: A survey on cyber vulnerabilities and defenses,”, Renewable and Sustainable Energy Reviews, Vol. 237, no. 117014, 2026.

[6] S. M. S. Hussain, T. S. Ustun, and A. Kalam, “A Review of IEC 62351 Security Mechanisms for IEC 61850 Message Exchanges,” IEEE Trans. Ind. Informatics, vol. 16, no. 9, pp. 5643–5654, 2020, doi: 10.1109/TII.2019.2956734

[7] S. M. Suhail Hussain, “Lightweight Optimized Message Authentication Scheme for IEC 61850 Sampled Value Messages,” IEEE Trans. Power Deliv., vol. 39, no. 4, pp. 2552–2555, 2024, doi: 10.1109/TPWRD.2024.3397551

[8] T. S. Ustun, S. M. S. Hussain, L. Yavuz, and A. Onen, “Artificial Intelligence Based Intrusion Detection System for IEC 61850 Sampled Values under Symmetric and Asymmetric Faults,” IEEE Access, vol. 9, pp. 56486–56495, 2021, doi: 10.1109/ACCESS.2021.3071141

[9] N. Cibin, B. Mulder, H. Carstens, P. Palensky, and A. Ştefanov, “Cyber Attacks Detection, Prevention, and Source Localization in Digital Substation Communication using Hybrid Statistical-Deep Learning,” *arXiv preprint* arXiv:2507.00522, 2026.

[10] D. Mishchenko, I. Oleinikova, and L. Erdodi, “Resilience of IEC 61850 Sampled Values-Based Protection Systems Under Coordinated False Data Injections,” *arXiv preprint* arXiv:2605.07535, 2026.

[11] A. Zaboli, S. L. Choi, T.-J. Song, and J. Hong, “A Novel Generative AI-Based Framework for Anomaly Detection in Multicast Messages in Smart Grid Communications,” *arXiv preprint* arXiv:2406.05472, 2024.

[12] F. Manzoor, V. Khattar, A. Herath, C. Black, M. C. Nielsen, J. Hong, C.-C. Liu, and M. Jin, “Detecting Zero-Day Attacks in Digital Substations via In-Context Learning,” *arXiv preprint* arXiv:2501.16453, 2025.

[13] Power Systems Management and Associated Information Exchange-Data and Communications Security-Part 6: Security for IEC 61850, Standard IEC 62351-6:2020.

[14] M. Rodriguez, J. Lazaro, U. Bidarte, J. Jimenez, and A. Astarloa, “A fixed-latency architecture to secure GOOSE and sampled value messages in substation systems,” IEEE Access, vol. 9, pp. 51646–51658, 2021, doi: 10.1109/ACCESS.2021.3069088

[15] J. Hong, R. Karnati, C. W. Ten, S. Lee, and S. Choi, “Implementation of Secure Sampled Value (SeSV) Messages in Substation Automation System,” IEEE Trans. Power Deliv., vol. 37, no. 1, pp. 405–414, 2022, doi: 10.1109/TPWRD.2021.3061205

[16] Information technology-Lightweight cryptography-Part 6: Message authentication codes (MACs), ISO/IEC 29192-6:2019.

[17] Power systems management and associated information exchange – Data and communications security – Part 9: Cybersecurity key management for power system equipment,” IEC 62351-9:2017.

[18] Weis, B.; Rowles, S.; Hardjono, T. The Group Domain of Interpretation. Internet Engineering Task Force (IETF) — Request for Comments (RFC) No. 6407. 2011. Available online: http://www.rfc-editor.org/info/rfc6407 (accessed on 02 April 2026).

[19] Maughan, D.; Schertler, M.; Schneider, M.; Turner, T. Internet Security Association and Key Management Protocol (ISAKMP). Internet Engineering Task Force (IETF) — Request for Comments (RFC) No. 2408. 1998. Available online: http://www.rfc-editor.org/info/rfc2408 (accessed on 02 April 2026).

[20] E. Rescorla, "The Transport Layer Security (TLS) Protocol Version 1.3," RFC 8446, Internet Engineering Task Force (IETF), Aug. 2018. Available online: https://datatracker.ietf.org/doc/html/rfc8446 ((accessed on 02 April 2026).

[21] N. M. Noel, V. O. Waziri, S. M. Abdulhamid et al., “Review and analysis of classical algorithms and hash-based post-quantum algorithm,” Journal of Reliable Intelligent Environments, vol. 8, pp. 397–414, 2022.

[22] J. Cuthrell “Smart Grids vs. Quantum Threats: The $1.45M Solution”, EEPOWER, July 2024. Available: https://eepower.com/tech-insights/smart-grids-vs-quantum-threats-the-1.45m-solution/

[23] K. Nakka, S. Ahmad, T. Kim, L. Atkinson and H. M. Ammari, "Post-Quantum Cryptography (PQC)-Grade IEEE 2030.5 for Quantum Secure Distributed Energy Resources Networks," 2024 IEEE Power & Energy Society Innovative Smart Grid Technologies Conference (ISGT), Washington, DC, USA, 2024, pp. 1-5

[24] J. Ahn, J. Chung, T. Kim, B. Ahn and J. Choi, "An Overview of Quantum Security for Distributed Energy Resources," 2021 IEEE 12th International Symposium on Power Electronics for Distributed Generation Systems (PEDG), Chicago, IL, USA, 2021, pp. 1-7.

[25] S.S.M. Reshikeshan, M.B. Koh and M.S. Illindala, "Rainbow Signature Scheme to Secure GOOSE Communications From Quantum Computer Attacks," in IEEE Transactions on Industry Applications, vol. 57, no. 5, pp. 4579-4586, Sept.-Oct. 2021.

[26] S. M. S. Hussain, M. A. Aftab, M. Zaery, S. M. Amrr, C. Konstantinou and M. A. Abido, "FALCON based Quantum Attack Secure Digital Signature Scheme for IEC 61850 GOOSE Messages," 2024 IEEE Industry Applications Society Annual Meeting (IAS), Phoenix, AZ, USA, 2024, pp. 1-6.

[27] W. Beullens, “Breaking Rainbow Takes a Weekend on a Laptop”, Cryptology ePrint Archive, 2022. https://eprint.iacr.org/2022/214

[28] “Module-Lattice-Based Key-Encapsulation Mechanism Standard,” NIST Rep., no. FIPS 203, p. FIPS 203, 2024, [Online]. Available: https://nvlpubs.nist.gov/nistpubs/FIPS/NIST.FIPS.203.pdf

[29] A. Ahmad and S. Jagatheswari, “Quantum Safe Multi-Factor User Authentication Protocol for Cloud-Assisted Medical IoT,” IEEE Access, vol. 13, no. November 2024, pp. 3532–3545, 2025, doi: 10.1109/ACCESS.2024.3523530.

[30] D. Dolev and A. C. Yao, “On the Security of Public Key Protocols,” IEEE Trans. Inf. Theory, vol. 29, no. 2, pp. 198–208, 1983, doi: 10.1109/TIT.1983.1056650.

[31] Open Quantum Safe, "liboqs Documentation," Available: https://openquantumsafe.org/liboqs/